\documentclass[aps,prd,showpacs,nofootinbib,superscriptaddress,preprintnumbers,twocolumn]{revtex4-1}
\usepackage{graphicx,amsmath,amssymb,soul,bbm,dcolumn,booktabs, multirow,array,float,enumitem,lmodern,dsfont,microtype,nicefrac}
\usepackage[dvipsnames]{xcolor}
\usepackage[utf8]{inputenc}
\newcolumntype{C}{>{$}c<{$}}
\AtBeginDocument{
\heavyrulewidth=.08em
\lightrulewidth=.05em
\cmidrulewidth=.03em
\belowrulesep=.65ex
\belowbottomsep=0pt
\aboverulesep=.4ex
\abovetopsep=0pt
\cmidrulesep=\doublerulesep
\cmidrulekern=.5em
\defaultaddspace=.5em
}
\newcommand{\be}{\begin{eqnarray}}
\newcommand{\ee}{\end{eqnarray}}

\def\fm {\mathop{\hbox{fm}}}
\def\MeV {\mathop{\hbox{MeV}}}

\def\Re {\mathop{\hbox{Re}}}

\def\beq{\begin{equation}}
\def\eeq{\end{equation}}
\def\beqs#1\eeqs{\beq\begin{split} #1 \end{split}\eeq}

\def\comment#1{}

\def\ket#1{\left| #1 \right\rangle}

\def\opbraket#1#2#3{ \left\langle #1 \left| #2 \right| #3 \right\rangle}
\def\av#1{ \left\langle #1 \right\rangle }

\usepackage{array}   
\newcolumntype{L}{>{$}l<{$}} 
\newcolumntype{S}{>{\footnotesize $}l<{$\normalsize}} 

\def\*#1{\mathbf{#1}}

\usepackage[colorlinks=true,backref=false, linktocpage=true,
citecolor=PineGreen,urlcolor=PineGreen,linkcolor=PineGreen,pdfpagemode=UseOutlines]{hyperref}
\hypersetup{%
  bookmarksnumbered=true,
  pdftitle = {},
  pdfsubject = {},
  pdfauthor = {},
  pdfkeywords = {}
}

\begin{document}
\title{Pion scattering in the isospin I=2 channel from elongated lattices}

\author{C.\ Culver}
\email{chrisculver@email.gwu.edu}
\affiliation{The George Washington University, Washington, DC 20052, USA}

\author{M.\ Mai}
\email{maximmai@gwu.edu}
\affiliation{The George Washington University, Washington, DC 20052, USA}

\author{A.\ Alexandru}
\email{aalexan@gwu.edu}
\affiliation{The George Washington University, Washington, DC 20052, USA}
\affiliation{Department of Physics, University of Maryland, College Park, MD 20742, USA}

\author{M.\ D\"oring}
\email{doring@gwu.edu}
\affiliation{The George Washington University, Washington, DC 20052, USA}
\affiliation{Thomas Jefferson National Accelerator Facility, Newport News, VA 23606, USA}

\author{F.\ X.\ Lee}
\email{fxlee@gwu.edu}
\affiliation{The George Washington University, Washington, DC 20052, USA}

\begin{abstract}
Pion-pion elastic scattering in the isospin $I=2$ channel is investigated in two-flavor dynamical lattice QCD.  Six ensembles are used with lattices elongated in one of the spatial dimensions
at two quark masses corresponding to a pion mass of 315 MeV and 226 MeV. 
The energy of the low-lying states below the inelastic threshold are extracted in each case using the standard variational method. 
The extracted finite-volume spectrum is fitted by the inverse amplitude method simultaneously for both 
quark masses and extrapolated thereafter to the physical point. The resulting phase-shifts and scattering length are compared with those from experiment, leading-order chiral perturbation theory and other lattice studies. 
Our calculations match the experimental results.
\end{abstract}

\pacs
{
12.38.Gc, 
14.40.-n, 
13.75.Lb  
}
\maketitle

\section{Introduction}\label{sec:intro}
The prediction of scattering phase-shifts of strongly interacting systems directly from quark-gluon dynamics has become possible through rapid advances in lattice QCD calculations. In such ab-initio calculations one can also vary parameters that are inaccessible in experiment like the number of flavors $N_f$. In addition, they serve to test models that extrapolate in hadron masses, allowing for deeper insights into the QCD dynamics at the hadronic scale, including resonances and their dependence on quark masses and flavors.  

Lattice QCD calculations are performed in a small cubic volume and in imaginary time, so that a direct evaluation of phase-shifts is not possible. However, the discrete energy eigenvalues of the QCD Hamiltonian in a cube with periodic boundary conditions can still be put into relation with phase-shifts as shown by L\"uscher~\cite{Luscher:1990ux, Luscher:1986pf}. Varying the lattice size, through, e.g., elongated boxes~\cite{Feng:2004ua, Lee:2017igf}, allows one to produce a number of phase-shifts at the eigenergies;  additional phase-shifts can be determined by projecting the particles to moving frames~\cite{Rummukainen:1995vs}. 
This enabled lattice QCD calculations of the $\rho$ meson phase-shifts in $N_f=2$~\cite{Aoki:2007rd, Feng:2010es, Gockeler:2008kc, Lang:2011mn, Pelissier:2012pi, Bali:2015gji, Guo:2016zos} and $N_f=2+1$~\cite{Aoki:2011yj, Dudek:2012xn, Feng:2014gba, Metivet:2014bga, Wilson:2015dqa, Andersen:2018mau, Fu:2016itp, Alexandrou:2017mpi}.
The isoscalar sector is particularly interesting because of the presence of the broad $f_0(500)$ ``$\sigma$'' resonance. Only recently, phase-shifts in this channel have been calculated~\cite{Briceno:2016mjc, Briceno:2017qmb, Guo:2018zss} (for calculations of the scattering lengths, see, e.g., Refs.~\cite{Fu:2017apw, Liu:2016cba}).

The infinite-volume extrapolation of the $\pi^+\pi^+$ system was the first physical application of the original L\"uscher formalism, for the scattering length~\cite{Sharpe:1992pp, Kuramashi:1993ka, Gupta:1993rn, Yamazaki:2004qb, Aoki:2005uf, Beane:2005rj, Beane:2007xs, Feng:2009ij, Yagi:2011jn, Fu:2013ffa,
Sasaki:2013vxa, Kawai:2017goq} and extended to higher energies~\cite{Dudek:2012gj, Bulava:2016mks, Akahoshi:2019klc}. See also Refs.~\cite{Helmes:2015gla, Helmes:2019dpy} for a comprehensive calculation of meson-meson scattering lengths. 
Systems of more than two pions with maximal isospin have been calculated in lattice QCD~\cite{Horz:2019rrn, Detmold:2008fn} and serve as a test ground for infinite-volume mapping techniques that are currently being developed~\cite{Blanton:2019igq, Pang:2019dfe, Briceno:2018aml, Mai:2018djl, Romero-Lopez:2018rcb, Guo:2018ibd, Hammer:2017kms, Briceno:2017tce, Hansen:2015zga, Detmold:2008gh}. In Ref.~\cite{Mai:2018djl} such a formalism was, for the first time, applied to the above-threshold $3\pi^+$ system.

In this study we determine the  $\pi^+\pi^+$ isospin $I=2$ phase-shifts using elongated boxes. 
Our study is carried out using $N_f= 2$ dynamical configurations with nHYP fermions~\cite{Hasenfratz:2007rf}. We analyze two sets of ensembles with different sea quark masses:  one corresponding to $m_\pi= 315$~MeV and the  other one to $m_\pi=  226$~MeV. For each pion  mass  we use three ensembles with different lattice geometry.  For each  ensemble  we  analyze  states  at  rest, ${\bf P}=  (0,0,0)$, and  states  moving  along  the  elongated  direction  with momentum ${\bf P}=  (0,0,1)$.   For  each  case,  we  use two-hadron $\pi\pi$ interpolators with different back-to-back momenta in the variational basis.  We extract 
the lowest states energy states using the variational method~\cite{Luscher:1990ck}.

The scattering length is computed at the two pion masses and the results are extrapolated to the physical point using chiral perturbation theory. In the wider energy range the inverse amplitude method (IAM)~\cite{Truong:1988zp, GomezNicola:2007qj} is utilized to fit the data and obtain predictions for the phase-shifts at the physical quark mass.
This model is unitary and matches the chiral pion-pion amplitude~\cite{Gasser:1984gg,Gasser:1983yg} up to the  next-to-leading order. The obtained predictions overlap nicely with the experimentally obtained values.

In comparison to previous studies in which this method was applied~\cite{Hu:2017wli, Doring:2016bdr, Hu:2016shf, Bolton:2015psa} we allow here the pion mass and decay constant to vary and include their full correlations with the energy eigenvalues in the fit.

In an upcoming paper~\cite{inprep} we will use the $I=2$ energy eigenvalues determined here, together with the corresponding results of the isovector and isoscalar channels~\cite{Guo:2016zos, Guo:2018zss}, to perform a global analysis of pion-pion-scattering with IAM. The full energy dependence of the $I=2$ partial-wave amplitude is also needed as input for upcoming calculations of the of $3\pi^+$ system above threshold along the lines of Ref.~\cite{Mai:2018djl}, as required by three-body unitarity~\cite{Mai:2017vot}. 

This paper is organized as follows. In Section~\ref{sec:lattice} details of the lattice calculation are provided, including the variational basis and interpolating operators, followed by a description of the extraction of energy eigenvalue in Section~\ref{sec:spectrum}. The determination of phase-shifts, effective range expansion fit, IAM fit, and
chiral extrapolation is discussed in Section~\ref{sec:physquant}.

\section{Lattice setup}\label{sec:lattice}


For the $I=2$ channel we cannot use $\bar{q}q$ interpolating fields since the maximum isospin for such operators is $1$. We will construct our variational basis out of two-pion interpolating fields, two $\pi^+$ pions to be precise, projected to the appropriate momentum combinations. Different interpolating fields will only differ by the choice of pion momenta. Using the interpolating fields we construct the correlator matrix:
\beq
    C_{ij}(t)=\av {\mathcal{O}_i(t)\mathcal{O}^{\dagger}_j(0)}.
\eeq
The eigenvalues of this matrix are obtained by solving the generalized eigenvalue problem,
\beq
C(t_0)^{-\frac{1}{2}}C(t)C(t_0)^{-\frac{1}{2}}\psi^{(n)}(t,t_0)=\lambda^{(n)}(t,t_0)\psi^{(n)}(t,t_0),
\eeq
where $t_0$ is a parameter.  The energies of the $\pi^+\pi^+$ system can be extracted from the long-time behavior of the eigenvalues,~\cite{Luscher:1990ck, Blossier:2009kd}
\beq
\lambda^{(n)}(t,t_0)\propto e^{-E_nt}\left[1 + \mathcal{O}(e^{-\Delta E_nt})\right], n=1,\ldots,N.
\eeq
When $t<2 t_0$ the correction term vanishes at a rate given by $\Delta E_n=E_{N+1}-E_n$, the difference between the  energy level of interest and the lowest level excluded from the variational basis~\cite{Blossier:2009kd}.  The variational method filters out contaminations in the spectrum from states included in the basis.

For each ensemble we include in the variational basis all interpolating fields with momenta corresponding to two-hadron states that in the absence of interactions will be below the inelastic thresholdat $E=4m_\pi$, and the next momentum just above it. In our analysis we include only the energies extracted from the variational method that lay below the inelastic threshold. 

In this work we will use two cubic lattices, and four lattices elongated in one spatial direction.  The relevant symmetry groups are $O_h$ and $D_{4h}$ for the cubic and elongated boxes, respectively.  The elongation direction is chosen to be $z$.  The angular momentum labels used for the irreducible representations (irreps) of $SO(3)$ are split into multiplets of the irreps of the lattice symmetry groups.  The splitting of angular momentum for $O_h$ and $D_{4h}$ can be found in Table~\ref{table:irep_split}.  

\begin{table}[t]
\begin{tabular}{r c c}
\toprule
$~~\ell~~~~~$ & $O_{h}$ & $D_{4h}$ \\
\hline
~~0~~~~~ & $A_1^+$                                        & $A_1^+$\\
~~1~~~~~ & $F_1^-$                                        & $A_2^-\oplus E^-$\\
~~2~~~~~ & $E^+ \oplus F_2^+$                             & $A_1^+\oplus B_1^+ \oplus B_2^+ \oplus E^+$\\
~~3~~~~~ & $A_2^- \oplus F_1^- \oplus F_2^-$              & $A_2^- \oplus B_1^- \oplus B_2^- \oplus 2 E^-$\\
~~4~~~~~ & $A_1^+ \oplus E^+ \oplus F_1^+ \oplus F_2^+$ & $2A_1^+ \oplus A_2^+ \oplus B_1^+ \oplus B_2^+ \oplus 2E^+$\\
\bottomrule
\end{tabular}
\caption{Resolution of angular momentum in terms of irreps of the $O_h$ and the $D_{4h}$ group.}
\label{table:irep_split}
\end{table}

To get additional states in this scattering region, we will also use interpolators for non-zero momentum states.  By constructing operators with a total momentum $\*P$ we will access additional energy levels in the elastic region.  The relativistic effects cause the box to shrink in the direction of the total momentum, changing the symmetry group.  We align the boost direction with the direction of elongation so that the relevant symmemtry group remains $D_{4h}$.  

The $\pi^+\pi^+$ interpolating fields are constructed using two $\pi^+$ interpolators with the appropriate momenta:
\beq
\pi\pi(\*P,\*p,t)\equiv\pi^+(\Gamma(\*p),t)\pi^+(\Gamma(\*P-\*p),t),
\eeq
where 
\beq
\pi^+(\Gamma(\*p),t)=\bar{d}(t)\Gamma(\*p)u(t).
\eeq
Here the quark fields $u(t)$ and $d(t)$ represent a three-dimensional slice
of the field, and they can be viewed as $N=4\times3\times V_3$ vectors,
where $V_3$ is the number of points in a time slice.
The matrix $\Gamma$ is an $N\times N$ matrix that acts in the spin, color and position space.  The only structure used in this study is $\Gamma(\*p)=\gamma_5 e^{i\*p}$. This matrix acts trivially in the
color space and in position space we have $[e^{i\* p}]_{\* x,\* y}=e^{i\* p \* x}\delta(\*x-\*y)$. For more details see Refs.~\cite{Pelissier:2012pi,Guo:2016zos}.

To access the zero angular momentum, $l=0$, phase-shifts in $\pi^+\pi^+$ we need to project our operators to the $A_1^+$ irrep of $O_h$ and $D_{4h}$. According to Table~\ref{table:irep_split} the lowest contribution to this irrep comes from $l=0$ states and the corrections come from $l=4$ states for $O_h$ and $l=2$ states for $D_{4h}$. These higher phase-shifts are expected to be small in the kinematic region we explore and can be safely neglected.

To construct interpolating fields with the right symmetry properties, we start with a ``seed'' interpolating field and project it on the $A_1^+$ irrep using:
\beq
    \pi\pi(\*P,\*p)=\frac{1}{|G|}\sum_{g\in G}\chi_{A_1^+}(g)\pi\pi(R(g)\*P,R(g)\*p).
    \label{eqn:op_project}
\eeq
Here $G$ is the group ($O_h$ or $D_{4h}$), $g$ is an element of $G$, $\chi$ is the character of $g$ in the irrep $A_1^+$ and $R(g)$ is the rotation corresponding to the element $g$.  

The collection of ``seed'' momenta used in this study are presented in Table~\ref{table:momcombos}.  As mentioned earlier, we used all momenta that in the non-interacting case had energy less than $4 m_\pi$ and the next just above. Then the operator is plugged into Eq.~(\ref{eqn:op_project}) to generate the linear combination that overlaps with $A_1^+$.  In Fig.~\ref{fig:energy_eta_315} we highlight which operators were used in the $315\MeV$ ensembles. For example, at elongation $\eta=1.25$ there are three energies in the elastic scattering region, thus four operators are selected.

\begin{table}[t]
\begin{ruledtabular}
\begin{tabular*}{0.99\columnwidth}{@{\kern1ex\extracolsep{\stretch{1}}}L*{8}{S}@{\kern1ex}}
                    && \multicolumn{3}{c}{$\* P=[000]$} &\kern 0.5em& \multicolumn{3}{c}{$\* P=[001]$} \\\midrule
\mathcal{E}_1^*     && [000]    & [001]    & [002]    && [001]    & [101]    & [002]    \\ \rule{0pt}{4ex} 
\multirow{2}{*}{$\mathcal{E}_2$} && [000]    & [001]    & [100]    && [001]    & [101]    & [002]    \\
                    && [101]    &          &          && [111]    &          &          \\\rule{0pt}{4ex} 
\multirow{2}{*}{$\mathcal{E}_3$} && [000]    & [001]    & [100]    && [001]    & [101]    & [002]    \\ 
                    && [101]    & [002]    & [102]    && [111]    & [003]    & [102]    \\ \specialrule{0.05em}{1ex}{1ex} 
\mathcal{E}_4^*     && [000]    & [001]    &          && [001]    & [101]    & [002]    \\\rule{0pt}{4ex}
\mathcal{E}_5                  && [000]    & [001]    & [100]    && [001]    & [101]    & [002]    \\\rule{0pt}{4ex}
\mathcal{E}_6                  && [000]    & [001]    & [100]    && [001]    & [101]    & [002]   
\end{tabular*}
    \end{ruledtabular}
\caption{The ``seed'' momentum $\*p$ used to create interpolating fields for the zero momentum states and the moving states. The momentum components are indicated in units of the smallest non-zero momentum allowed in the corresponding direction; for the elongated boxes the smallest momentum in the $z$-direction is reduced proportional to the elongation.  Each of these operators is projected onto the $A_1^+$ irrep for the $D_{4h}$ group in the elongated case, or $O_h$ group for the cubic case (indicated with an star in the enesemble label).}
\label{table:momcombos}
\end{table}

These operators are evaluated on a set of six ensembles. The parameters for these ensembles are listed in Table~\ref{table:gwu_lattice}. To compute the correlation functions we need to perform the Wick contractions and the correlation functions become functions of quark propagators. All correlation functions used in this study are linear
combinations of 
\beqs
C(\*P,&\*p,\*p',t)\equiv \av{\pi\pi(\*P,\*p',t)\pi\pi(\*P,\*p,0)^\dagger}\\
=&-[5(\*P-\*p')t\mid 5(-\*P+\*p)0\mid 5\*p' t\mid 5(-\*p)0]\\
&+[5(\*P-\*p')t\mid 5(-\*P+\*p)0]\,[5\*p't\mid 5(-\*p)0]\\
&-[5(\*P-\*p')t\mid 5(-\*p)0\mid 5(\*p')t\mid 5(-\*P+\*p)0]\\
&+[5(\*P-\*p')t\mid 5(-\*p)0]\,[5(\*p')t\mid 5(-\*P+\*p)0].
\eeqs
Above we introduced the following notation to denote the quark
propagator traces:
\beq
    \left[i_1\*p_1j_1\mid\ldots\mid i_k\*p_kj_k\right]\equiv\text{Tr}\prod_{\alpha=1}^{k}\Gamma(\*p_{\alpha})M^{-1}(t_{j_{\alpha}},t_{j_{\alpha+1}})\,,
    \label{eqn:diagram_trace}
\eeq
where $M^{-1}(t,t')$ represents the quark propagator $N\times N$ matrix between two time-slices.


Evaluating these diagrams requires the all-to-all quark propagator. This is a very expensive calculation and to avoid it we use the Laplacian-Heaviside method~(LapH)~\cite{Peardon:2009gh}. The basic idea is to replace the quark interpolating fields with smeared quark interpolators that have the same symmetries. The smearing is introduced by truncating the three-dimensional Laplacian operator on each time-slice by keeping the lowest-lying $N_v$ operator modes. The net effect is that the point-quark all-to-all propagator $M^{-1}$ is replaced with the propagator for smeared quarks ${\widetilde M}^{-1}$. The smeared propagator can be computed efficiently since we only have to invert the Dirac matrix for the LapH modes, rather than for each point on the lattice. We stress that this replacement is not an approximation, but rather creates operators with the right quantum numbers which have a different overlap with the relevant states. The number of LapH modes $N_v$ controls the smearing radius and the overlap; if the truncation is aggressive, keeping only a few LapH modes, the overlap decreases and correlation functions need to be computed accurately at large time separations to resolve the relevant states. For our study we used $N_v=100$ which corresponds to a smearing radius of roughly $0.5\fm$~\cite{Guo:2016zos}. To compute the smeared propagator efficiently we use GPU inverters~\cite{Alexandru:2011ee}.
In our calculations all steps remain unchanged except that the quark traces in Eq.~(\ref{eqn:diagram_trace}) are computed using the smeared propagator ${\widetilde M}^{-1}$.

\begin{table*}[t]                   

\begin{ruledtabular}

\begin{tabular}{@{}*{13}{>{$}l<{$}}@{}}                                                                
\text{ensemble}~& ~N_t\times N_{x,y}^2\times N_z~ & ~\eta~ & ~a[\fm]~   & ~N_\text{cfg}~  & ~aM_{\pi}~  & ~am^{pcac}_{u/d}~  & ~af_{\pi}~  \\
\midrule                                                                                               
\mathcal{E}_1&48\times24^2\times24  &  1.00  & 0.1210(2)(24) & 300   & 0.1931(4)  & 0.01226(5) & 0.0648(8)~    \\ 
\mathcal{E}_2&48\times24^2\times30  &  1.25 & -        & -     &    0.1944(3)      &     0.01239(4)     &  0.0651(6)~      \\
\mathcal{E}_3&48\times24^2\times48  &  2.00  & -        & -     &    0.1932(3)   &     0.01227(5)     &  0.0663(6)~         \\
\mathcal{E}_4&64\times24^2\times24  &  1.00  & 0.1215(3)(24)& 400  & 0.1378(6)  & 0.00612(5)~& 
0.0600(10)  \\                                                                                           
\mathcal{E}_5&64\times24^2\times28  &  1.17 &      -   & 378     & 0.1374(5) &   0.00620(4)       &   0.0600(8)~         \\
\mathcal{E}_6&64\times24^2\times32 &  1.33 &      -   & 400     & 0.1380(5)      &  0.00619(4)      &   0.0599(10)      \\
\end{tabular}  
\end{ruledtabular}
\caption{A summary of lattice details used including the lattice spacing $a$, number of gauge configurations.  The nucleon mass, pion decay constant and kaon decay constant are represented.}
\label{table:gwu_lattice}                                                                                  
\end{table*}      

\section{Extracting finite-volume spectrum}\label{sec:spectrum}

\begin{figure}[t]
    \centering
    \includegraphics[width=\columnwidth, trim=0.2cm 0.5cm 0.75cm 0.8cm]{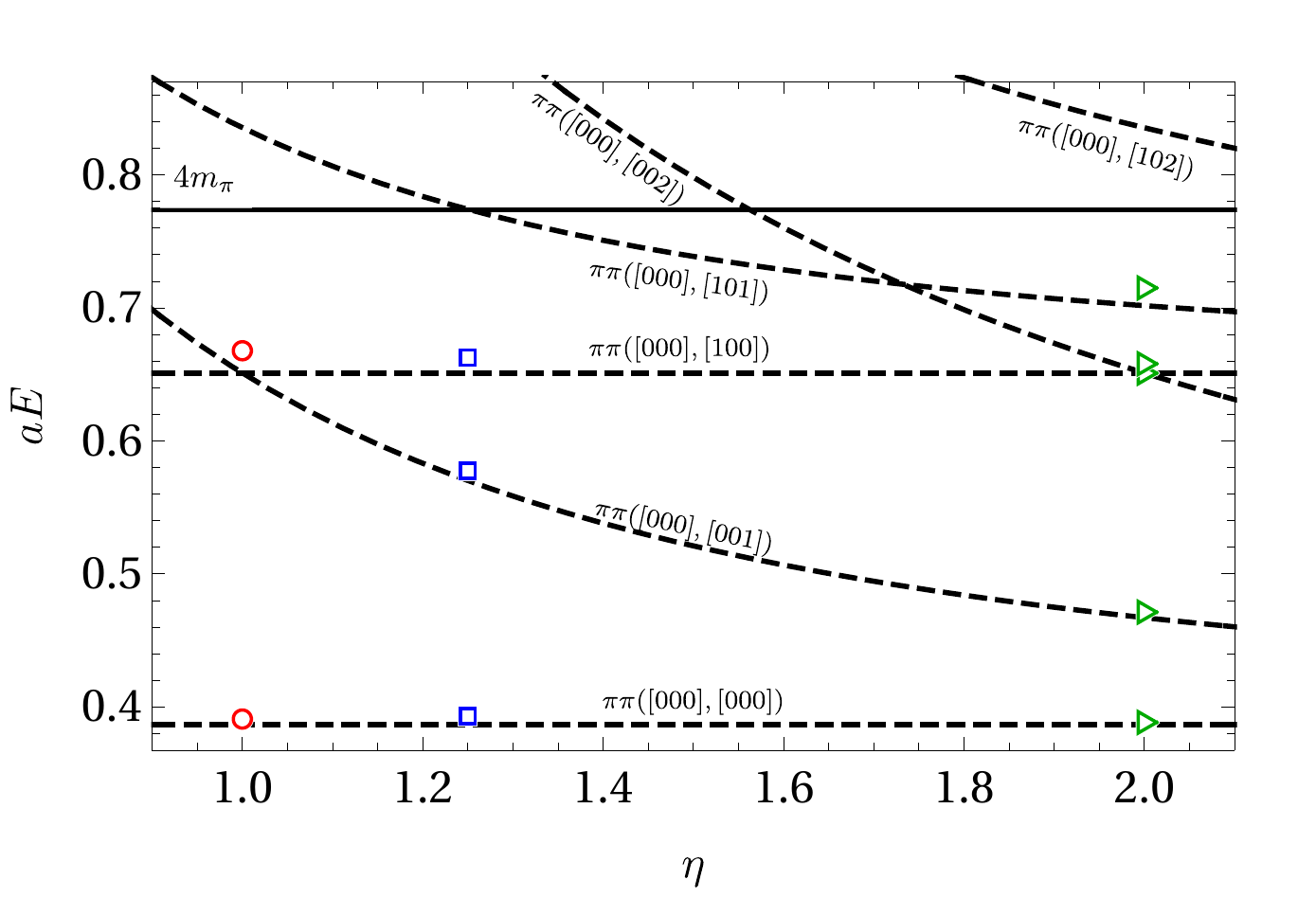}
    \caption{Energy levels extracted from the $315\MeV$ ensembles with total momentum $\*P=[000]$ as a function of elongation.  The dashed lines correspond to the non-interacting $\pi\pi$ energies.  The lattice data are plotted with the error bars within the markers.
    In the figure the notation follows the text: the $\pi\pi(\*P,\*p)$ interpolator has total momentum $\*P$
    and relative mometum $\*p$.}
    \label{fig:energy_eta_315}
\end{figure}

\begin{figure}[t]
    \centering
    \includegraphics[width=\columnwidth, trim=0.2cm 0 0.8cm 0.5cm]{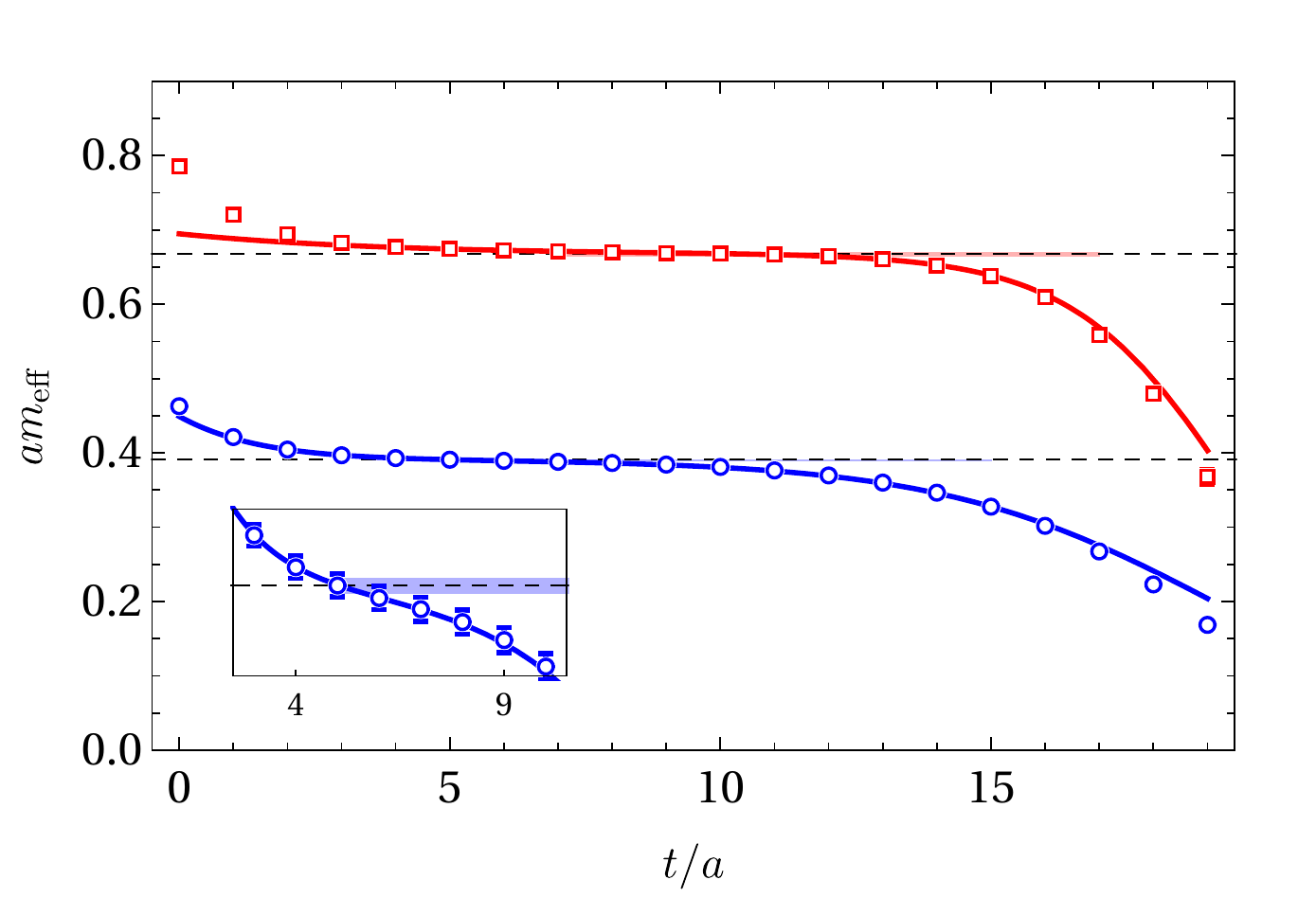}
    \caption{Effective masses for the two lowest $\*P=[000]$ levels in the ${\cal E}_1$ ensemble. The solid line corresponds to the fitted function. The dashed lines correspond to the energies extracted from the fit, the thin barely visible rectangles correspond to the error-bands and they span the fitting range. The error bars are present but for most points smaller than the symbol size. In this plot $t$ is measured from $t_0$.}
    \label{fig:effm}
\end{figure}

The variational method gives us the generalized eigenvalues as a function of time separation between 
sink and source. To extract the energy levels we have to fit these functions. When the correlation functions
are saturated by a single state, the mass can be extracted using a single-exponential fit. To determine
the appropriate range for this fit we plot the effective mass, $m_\text{eff}\equiv -\log \lambda(t+1)/\lambda(t)$,
and look for a plateau. As we can see from Fig.~\ref{fig:effm}, the data is very precise and the effect of
the corrections is statistically significant for all time ranges~(see the inset). As such, we have to fit the correlators
to include the effects of the excited states and thermal corrections:
\beq
\lambda(t) = A_1 e^{-E_1 (t-t_0)} + A_1' e^{-E_1'(t-t_0)}+ B e^{-\Delta E (t-t_0)} \,.
\eeq
Above the fitting parameters are the energy levels $E_1$ and $E_1'$, the spectral weights $A_1$ and $A_1'$, and
$B$ the coefficient of the leading thermal correction, to be discussed below. The inclusion of the excited level allows us to
fit the correlator at earlier times and $E'_1$ should have a value corresponding to the lowest states excluded from the variational basis. 
This is indeed the case in our analysis.

One unusual feature for two-pion correlation functions is that the wrap-around effects, due to thermal fluctuations,
lead to constants or slowly decaying exponential contributions to the correlation function~\cite{Dudek:2012gj}.
To see how the thermal effects contribute, consider the correlation function 
\beq
    C(t)=\frac1Z\sum_{n}\opbraket{n}{e^{-H(T-t)}\pi\pi(\*P,\*p') e^{-Ht}\pi\pi(\*P,\*p)^\dagger}{n}\,,
\eeq
where $T$ is the time-extent of the lattice. The leading order contribution comes from $\ket{n}=\ket{0}$, which leads to the usual superposition of exponentials for the correlation function. The next order term for $\pi\pi$ comes from $\ket{n}=\ket{\pi(\*p)}$, single pion states with momentum $\*p$. The effect of these states is to generate a set of slowly decaying exponentials
\beq
\delta C(t) = \frac1Z \sum_{\*p} e^{-E_\pi(\*p)T} e^{-\Delta E(\*p) t}\,, 
\eeq
with $E_\pi(\*p)$ the energy of a single pion with momentum $\*p$ and $\Delta E=E_\pi(\*P+\*p)-E_\pi(\*p)$. Note that these contributions are suppressed by coefficients that vanish exponentially fast with $T$. However, when fitting our correlators for $t\sim T/2$, their effects are important, as can be seen from Fig.~\ref{fig:effm}, and this correction needs to be included. In our fits we fix the exponential term $\Delta E$ to the lowest value generated by varying $\*p$ over the allowed momenta: for $\*P=0$ we fix $\Delta E=0$ and
for $\*P=[001]$ we fix $\Delta E=\sqrt{m_\pi^2+(2\pi/L/\eta)^2}-m_\pi$.

\begin{figure}[t]
    \centering
    \includegraphics[width=\columnwidth, trim=0.5cm 1.cm 1.9cm 2cm]{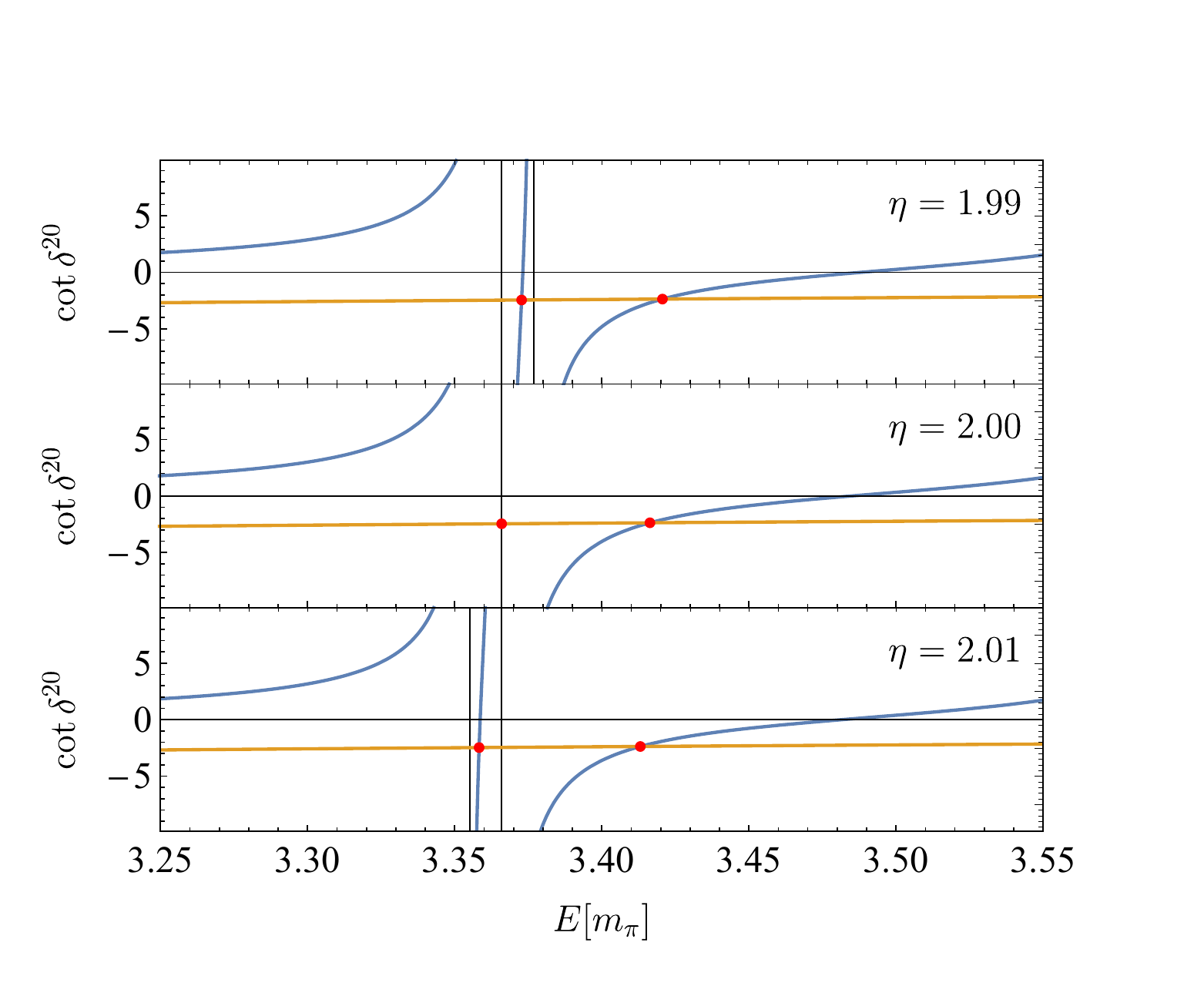}
    \caption{Energy levels as derived from Eq.~\eqref{eq:phase-shift} at the intersection of the phase-shift curve generated by the IAM model (orange line) and ${\cal W}_{00}$ (blue line.) The vertical lines indicate the position of the non-interacting energies. The three panels correspond to elongations around $\eta=2$, to help visualize how the interacting energy is forced to coincide with the non-interacting level.}
    \label{fig:pinch-pole}
\end{figure}

For elongated boxes an interesting feature arises: the non-interacting levels cross as we vary the elongation. This can be easily
seen in Fig.~\ref{fig:energy_eta_315}. In particular, we draw the reader's attention to the intersection between the
$\pi\pi([000],[100])$ and $\pi\pi([000],[002])$ that occurs at $\eta=2$. This has consequences also for the energy levels
in the interacting case. {\em Even in the presence of interactions, states appear with energies at (or very close to) non-interacting levels}. To understand this note that the interacting energy levels fall in between two non-interacting ones. This can be understood
in the case that the only partial wave contributing is $\ell=0$. In this case the energy levels satisfy Eq.~(\ref{eq:phase-shift}) below
and the ${\cal Z}_{00}$ function has poles at the energies corresponding to non-interacting levels, bracketing the interacting
solutions.  In Fig.~\ref{fig:pinch-pole} we show the solutions of Eq.~(\ref{eq:phase-shift}) as we vary the elongation of the box past $\eta=2$. Notice the vertical lines that are the poles of the ${\cal Z}_{00}$ functions: one corresponding to $\pi\pi([000],[100])$ state that remains fixed since the momentum in the transverse directions does not change as we change elongation, and one corresponding to $\pi([000],[002])$ that moves as we vary $\eta$. The two lines bracket one of the solutions of Eq.~(\ref{eq:phase-shift}) and when $\eta=2$ the poles merge and the corresponding solution has exactly the same energy as the non-interacting two-pion state. 

Note that this argument relies on Eq.~(\ref{eq:phase-shift}) being exact, which only works in the limit where the higher partial waves are zero. We expect that when the other partial waves are taken into account, the energies will be shifted slightly away from this position. This is for example the case when the poles for $\pi\pi([000],[100])$ and $\pi\pi([000],[001])$ merge, which happens for $\eta=1$, corresponding to a cubic box. In that case, the energy level that is pinched by the poles ends up belonging to the $E^+$ representation of the symmetry group for cubic boxes $O_h$. The lowest angular momentum that this irrep overlaps with is $\ell=2$. In this case it is then natural that the energy for this level coincides with the non-interacting case if we assume that all partial waves above $\ell=0$ are small: the shift will be proportional to the $\delta^{I=2,l=2}$. 

It is important to emphasize that the presence of these levels do not imply that the phase-shift $\delta^{20}$ is zero for this energy, as one would expect. For example in Fig.~\ref{fig:pinch-pole} the phase-shift is clearly non-zero at the pole. To understand this note that the connection between the phase-shift and energy is controlled by the ${\cal Z}_{00}$ function which is infinitely steep when the poles merge. Thus, finite changes in the phase-shift away from zero lead to infinitesimal (zero) changes in the energy away from the pole. This is also important when analyzing the levels extracted from numerical simulations. The energy levels will be determined with some finite stochastic error, which is mapped through this infinitely steep function into infinite errors in the phase-shift space. As such, these energy levels offer no real constrain on the phase-shifts and we do not include them in our analysis. Note that large error bars in phase-shifts always arise when the error-bars cross a pole, which happens even without pinching (see for example the left panel of Fig.~\ref{fig:I2phase-shifts}).

\section{Extracting physical quantities}\label{sec:physquant}

Having computed the finite-volume energy levels of $\pi\pi$ scattering in maximal isospin, we now connect them to physical quantities.  For energies below the inelastic threshold we use L\"uscher's formula~\cite{Luscher:1990ux} and the extensions to elongated boxes~\cite{Feng:2004ua} and boosted frames~\cite{Lee:2017igf}.  The s-wave is the lowest partial wave that contributes to $\pi\pi$ scattering in isospin 2.  For cubic boxes, elongated boxes, and boosted systems, the irrep that overlaps with $\delta^{20}$ is $A_1^+$.  Higher partial waves ($l=2,4,6\ldots$) also overlap with this irrep in a finite volume, however, we neglect these from further analysis since they are known to be negligible in this channel, see, e.g., Refs.~\cite{Pelaez:2015qba,Dudek:2012gj}.  Thus, for the extraction of the phase-shifts in this channel the relevant L\"uscher formula is
\beq
    \cot\delta^{20}=\mathcal{W}_{00}=\frac{\mathcal{Z}_{00}(1,q^2;\eta)}{\pi^{3/2}\eta q}\,.  
    \label{eq:phase-shift}
\eeq
Boosting the system has the effect of changing the value of $\eta$, see Ref.~\cite{Lee:2017igf} for further details and explicit form of $\mathcal{Z}_{00}(1,q^2;\eta)$. In the following we will use two distinct parameterizations of the phase-shifts, which also will allow us to interpolate them in energy as well as to extrapolate them to the physical point. 

\subsection{Effective range expansion}\label{sec:ERE}

The effective range expansion (ERE) is expected to hold in channels without resonances in vicinity of the production threshold. The drawback is that with a finite number of terms we can only describe low momentum data. We will use the first two terms of the ERE, i.e.,
\beq
    p\cot\delta(p)=\frac{1}{a_0}+\frac{1}{2}r_0p^2\,.
\eeq
The parameters $a_0$ and $r_0$ are the scattering length and effective range, respectively.  The lattice energy spectrum can be used to compute phase-shifts, and we can find an $a_0$ and $r_0$ from a fit to this results.  To perform this fit, the above-given phase-shift is related to the finite-volume spectrum via the L\"uscher's formula.  The corresponding correlated $\chi^2$ with the energy eigenvalues is minimized then with respect to $a_0$ and $r_0$.

For both pion masses we restrict the fit to the lowest two energy levels in each ensemble.  For the $315\MeV$ data the scattering length extracted is $m_{\pi}a_0=-0.20(3)$ with a $\chi^2/\text{dof}=5.6/(6-2)$.  The $226\MeV$ ensemble yields a scattering length $m_{\pi}a_0=-0.10(2)$ with a $\chi^2/\text{dof}=3.3/(6-2)$. 
The $315\MeV$ scattering length is within two sigma of the leading order (LO) ChPT value for that pion mass, while the $226\MeV$ scattering length is consistent with the LO ChPT value within one sigma.

\begin{figure}[t]
    \centering
    \includegraphics[width=\columnwidth,trim= 0.2cm 1cm 0 0]{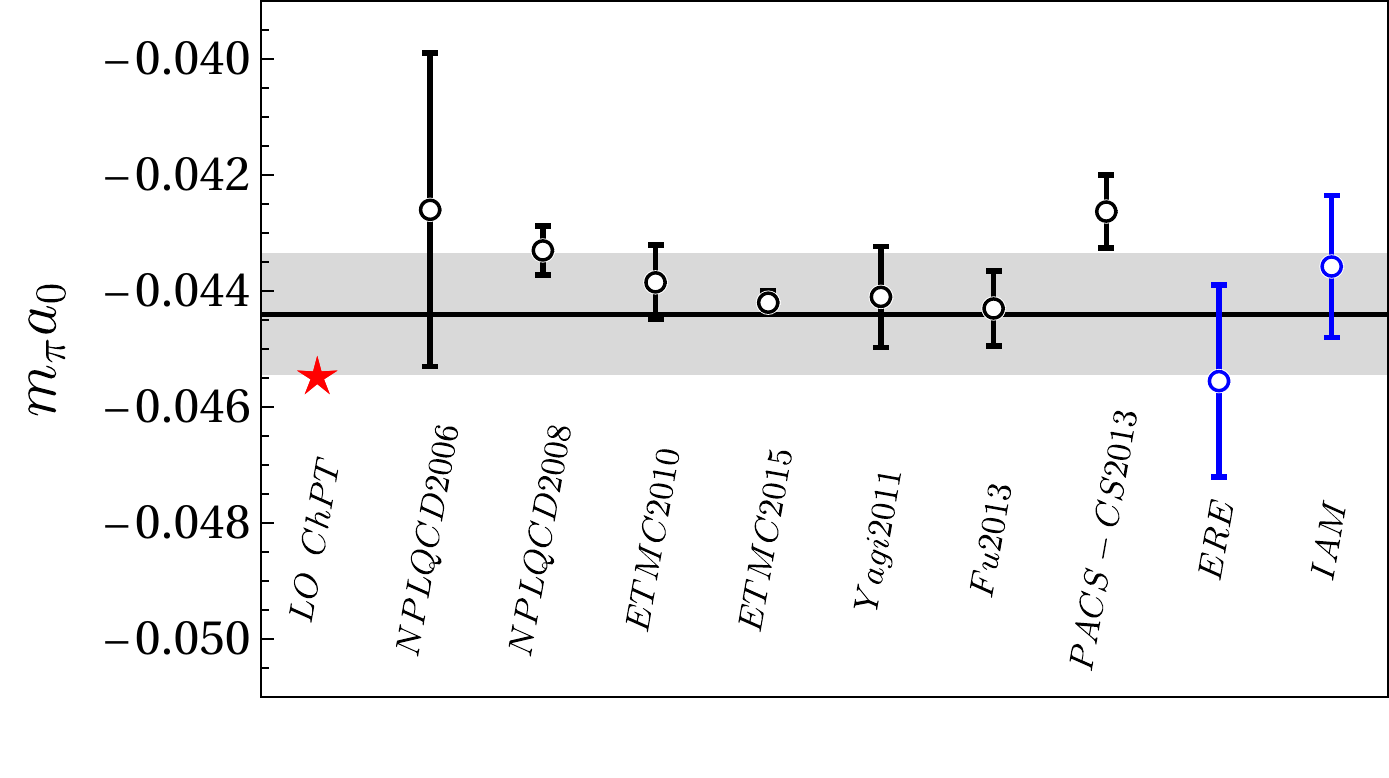}
    \caption{Plot of the scattering length as extracted from this, ``ERE'' and ``IAM,'' and other lattice QCD studies in a similar representation as in Ref.~\cite{Helmes:2019dpy}.  The red star is the LO ChPT value for the scattering length~\cite{Gasser:1983yg} and the gray band depicts the more recent result using Roy equations~\cite{Ananthanarayan:2000ht}.
   For the other references, see main text.
    }
    \label{fig:scattering_lengths}
\end{figure}

We can extrapolate these scattering lengths to the physical pion mass using ChPT at NLO~\cite{Gasser:1983yg}.  The expansion of $m_{\pi}a_0$ reads
\beq
    m_{\pi}a_0=-\frac{m_{\pi}^2}{16\pi f_{\pi}^2}\left[1+\frac{m_{\pi}^2}{32\pi^2f_{\pi}^2}\left(3\ln\frac{m_{\pi}^2}{2f_{\pi}^2}-1-l_{\pi\pi}\right)\right]\,,
\eeq
where $l_{\pi\pi}$ is a combination of the usual low-energy constants (LECs). This function is fit to our two extracted scattering lengths at unphysical pion masses, with respect to $l_{\pi\pi}$. With a $\chi^2/\text{dof}=0.74/(2-1)$ we obtain a value of $l_{\pi\pi}=-1.09(2.52)$.  Evaluating the ChPT expansion at the physical pion mass we obtain $m_{\pi}a_0=-0.0455(16)$.  We compare this result to the physical value and other lattice studies in Fig.~\ref{fig:scattering_lengths} indicated as ``ERE.''  The gray band represents the result of the estimation from the Roy equation~\cite{Ananthanarayan:2000ht}.  
The references for other determinations of the scattering length are, in order: NPLQCD2006~\cite{Beane:2005rj}, NPLQCD2008~\cite{Beane:2007xs}, ETMC2010~\cite{Feng:2009ij}, ETMC2015~\cite{Helmes:2019dpy}, Yagi2011~\cite{Yagi:2011jn}, Fu2013~\cite{Fu:2013ffa}, and PACS-CS2013~\cite{Sasaki:2013vxa}. This figure also contains a second scattering length from the lattice data using the inverse amplitude method as explained below, indicated as ``IAM.''

\subsection{Inverse amplitude method}\label{sec:IAM}

\begin{figure*}[t]
    \centering
    \begin{tabular}{ccc}
    \parbox[c]{5.55cm}{\includegraphics[width=\linewidth,trim=1.1cm 0 2.2cm 0]{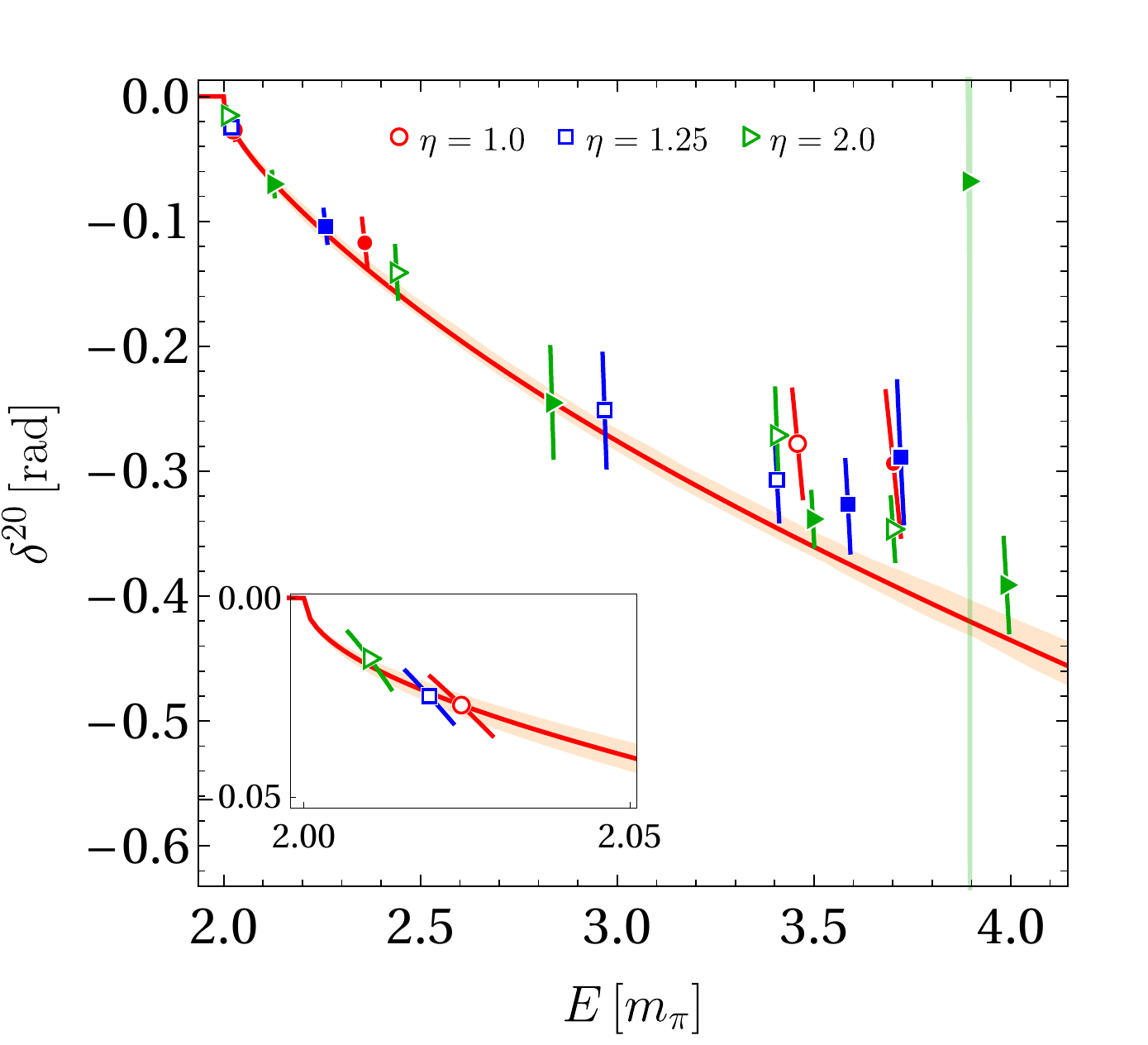} }&
    \parbox[c]{5.5cm}{\includegraphics[width=\linewidth,trim=1.1cm 0 2.2cm 0]{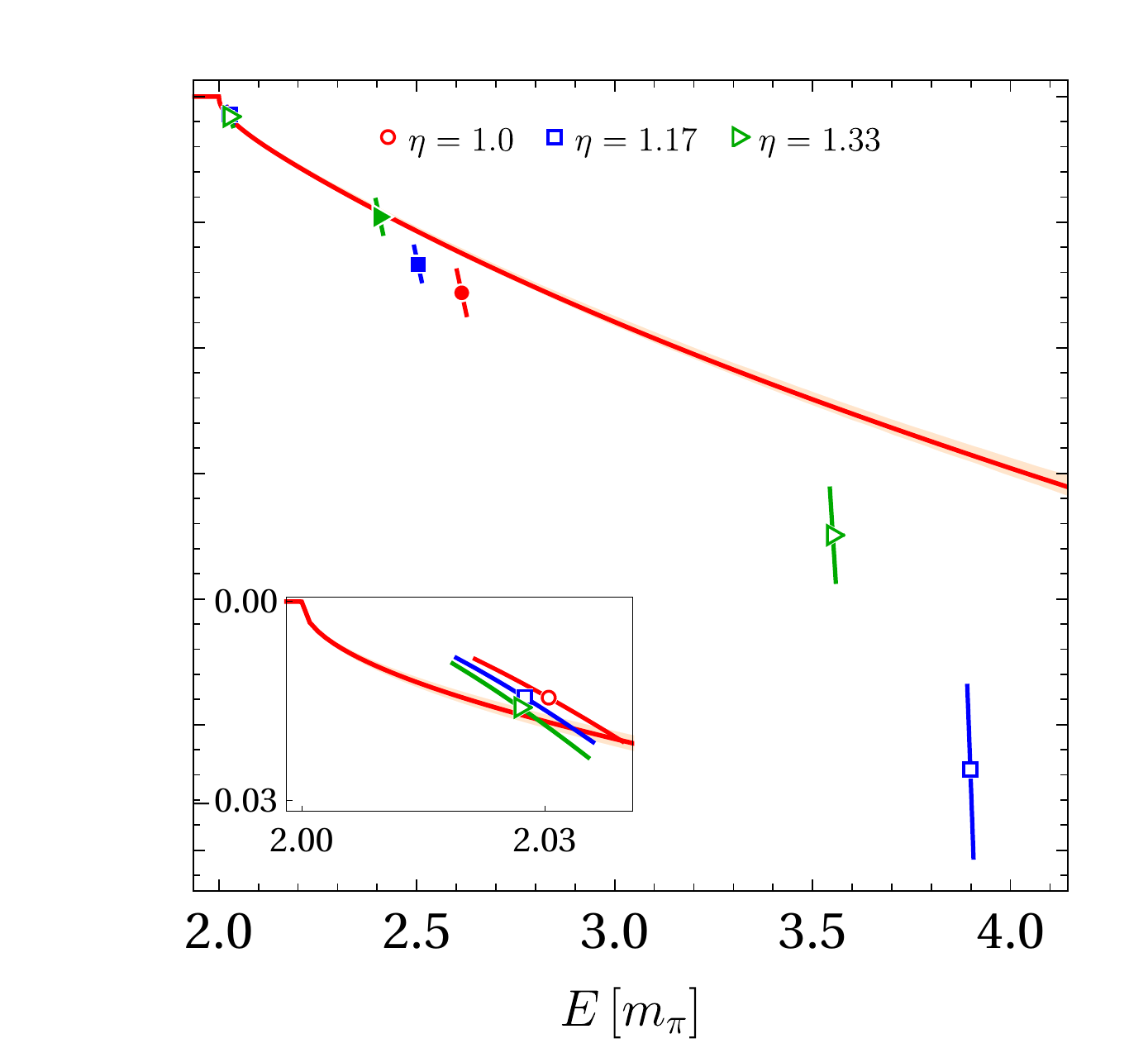} }&
    \parbox[c]{5.5cm}{\includegraphics[width=\linewidth,trim=1.1cm 0 2.2cm 0]{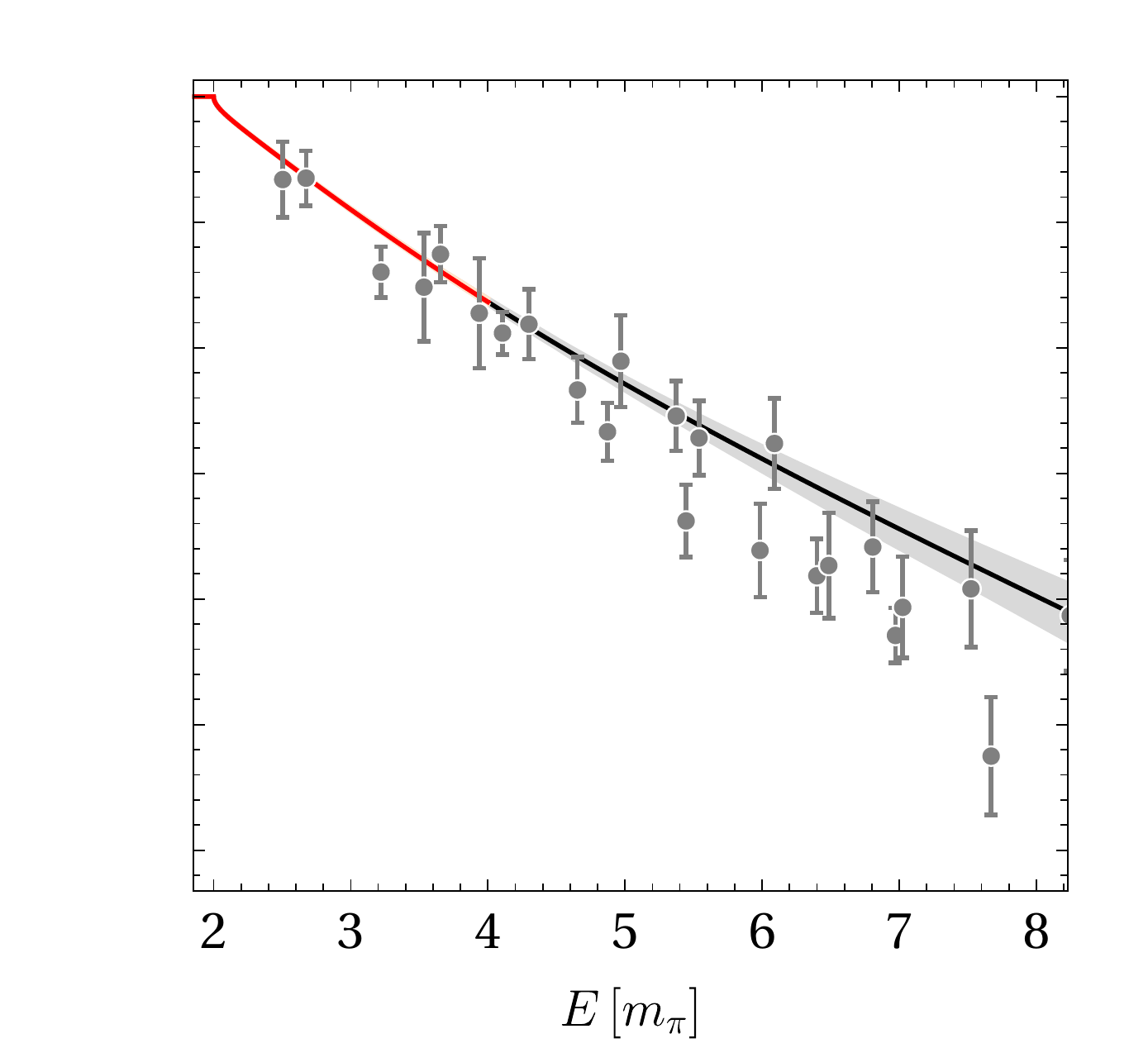} }
    \end{tabular}
    \caption{Plots of the phase-shifts as a function of energy for pion masses $315\MeV,226\MeV,139\MeV$ in order from left to right.  The points on the first two plots are lattice QCD data as described in the text.  The empty points are for systems with $\*P=[000]$ while the filled points are for systems with $\*P=[001]$.  The phase shifts extracted from experiment (right panel) are from Refs.~\cite{Rosselet:1976pu, Estabrooks:1974vu}.  The curves and error bands show the result of an IAM fit to the $I=2$ lattice QCD energy eigenvalues and the pertinent prediction at the physical point in the right panel.}
    \label{fig:I2phase-shifts}
\end{figure*}

The available data on the finite-volume spectrum of $\pi\pi$ scattering at maximal isospin covers a large region in energy and pion mass. Therefore, to connect those data to each other but also to the physical point one requires a framework that can extrapolate in energy and $m_\pi$. The so-called Inverse Amplitude Method (IAM)~\cite{Truong:1988zp,GomezNicola:2007qj,GomezNicola:2007qj} reconciles both these requirements. In particular, it provides a scattering amplitude, which fulfills two-body unitarity exactly and has correct chiral behaviour up to next-to-leading chiral order~\cite{Gasser:1984gg,Gasser:1983yg}. Furthermore, it fulfills the general requirements on the chiral trajectory for resonances, derived in Ref.~\cite{Bruns:2017gix} to all chiral orders. The corresponding cotangent of the phase-shift reads
\begin{align}
\cot\,\delta_{\rm IAM}(s)=
\frac{\sqrt{s}}{2p}
\Bigg(\frac{T_2(s)-\Re{T_4(s)}}{(T_2(s))^2}\Bigg)\,,
\label{eq:IAM}
\end{align}
where $T_2(s)$ and $T_4(s)$ are the leading and next-to-leading order chiral amplitudes, respectively, projected to isospin $I=2$ and angular momentum $l=0$. The total energy squared of the system is denoted by the Mandelstam $s=E^2$.

The leading order chiral amplitude $T_2(s)$ is a function of energy, Goldstone-boson mass, $m^2=B(m_u+m_d)$ and pion decay constant in the chiral limit, $f_0$. The amplitude $T_4$ involves in the two-flavor case two low-energy constants (LECs) $\bar l_1$ and $\bar l_2$. Two additional low-energy constants $\bar l_3$, $\bar l_4$ enter the NLO chiral amplitude when replacing the above mass and decay constants by their physical (lattice) values using one-loop results~\cite{Gasser:1983yg},
\begin{align*}
m_\pi^2=m^2\left(1-\frac{m^2}{32\pi^2f_0^2}\bar l_3\right)
\text{~\&~}
f_\pi=f_0\left(1+\frac{m^2}{16\pi^2f_0^2}\bar l_4\right)\,.
\end{align*}
The constants $\bar l_i$ do not depend on the regularization scale, but only on the parameters of the underlying theory - the quark masses. However, they are related to the scale-dependent, but quark-mass independent renormalized LECs via
\begin{align*}
l_i^r=\frac{\gamma_i}{32\pi^2}\left(\bar l_i+\log \frac{m^2}{\mu^2}\right)\,,
\end{align*}
where $\gamma_1=\nicefrac{1}{3}$, $\gamma_2=\nicefrac{2}{3}$, $\gamma_3=-\nicefrac{1}{2}$, $\gamma_4=2$. For a fixed scale $\mu$ one can, thus, determine the renormalized LECs and then make predictions for the two-particle scattering for a setup with a different pion mass. In the course of this work we use dimensional regularization with $\mu=770$~MeV, but emphasize that the expression~\eqref{eq:IAM} is manifestly scale independent. 

As discussed before, the finite volume spectrum consists of a large set of  energy eigenvalues as well as $m_\pi$ and $f_\pi$. In the past it has been  noted~\cite{Bulava:2016mks} that variations of the pion mass and decay constant can lead to non-negligible effects in the energy spectrum. Therefore, it is important to asses this source of uncertainty in a systematic way. In relation to this we noted some systematic  effects in $\mathcal{E}_2$, c.f., different central value of the pion mass recorded in Table~\ref{table:gwu_lattice}. Thus, we exclude this set from further fits, which leaves us with 21 energy eigenvalues, five pion masses and five decay constants to fit. The latter is performed plugging in Eq.~\eqref{eq:IAM} into Eq.~\eqref{eq:phase-shift} while minimizing the correlated $\chi^2$ with respect to 3 dynamical parameters~\footnote{We found that the value of $l_r^3$ does not lead to any notable improvement of the fit and fix this value to the one reported by FLAG~\cite{Aoki:2019cca}: $l_r^3=8.94\cdot10^{-6}$.} $l_r^1$, $l_r^2$ and $l_r^4$. We let the fit also determine the  values of the pion mass and decay constants that are common for the three light and three heavy ensembles, respectively. Note that all correlations between pion mass, decay constant and energy eigenvalues are taken into account. The best parameters of the overall fit to light and heavy data produce $\chi^2/{\rm dof}=75.4/(31-7)$ with the LECs given in Table~\ref{tab:lec}.

\begin{table}[h]
    \centering
    \begin{ruledtabular}
   \begin{tabular}{c|c}
    \addlinespace[0.2em]
    \multicolumn{2}{c}{
    ~~~~$l_r^1=+11.6^{+16.2}_{-14.4}$~~~~
    $l_r^2=-0.7^{+6.9}_{-7.4}$~~~~
    $l_r^4=+52.4^{+25.1}_{-25.5}$~~~~
    }
    \\[0.4em]
    \hline
    \addlinespace[0.4em]
    ~~~$m_\pi^{\rm light}=223.83^{+0.19}_{-0.18}$~MeV&
    $m_\pi^{\rm heavy}=315.25^{+0.06}_{-0.09}$~MeV~~~\\[+0.2cm]
    ~~~$f_\pi^{\rm light}=97.45^{+0.56}_{-0.30}$~MeV&  
    $f_\pi^{\rm heavy}=107.44^{+0.16}_{-0.25}$~MeV~~~\\[0.2em]
    \end{tabular}
    \end{ruledtabular}
    \caption{The fitted LECs ($l_r^i\cdot 10^3$) and $m_\pi$ and $f_\pi$ from the IAM analysis.}
    \label{tab:lec}
\end{table}

\noindent
The errors on the parameter have been estimated in a re-sampling procedure. We have explored the origin of the large $\chi^2$ of this combined fit extensively. As a matter of fact, the fit to the results of only heavy or light ensembles gives $\chi^2/{\rm dof,heavy}=21.2/(17-5)$ and $\chi^2/{\rm dof,light}=7.0/(14-5)$.
Thus, it appears that IAM is flexible enough to address all lattice QCD results up to the inelastic threshold as a function of energy. However, the simultaneous parametrization of the pion mass dependence is less reliable for such a large pion mass range. To some extent this is expected, since IAM coincides with the chiral expansion only up to next-to-leading chiral order. Furthermore, we refer at this point to the recent study of systematic uncertainties tied to the use of lattice data at unphysical pion mass in the context of ChPT~\cite{Durr:2014oba}.

We note that, for the individual fits where the model fits well the data,
a significant part of the $\chi^2$ comes from the cross-correlation between the data points. This demonstrates that the cross-correlations should be taken into account.

Table~\ref{tab:a0} lists the scattering lengths for the obtained set of parameters.

\begin{table}[h]
\centering
\begin{ruledtabular}
\begin{tabular}{cccc}
~&heavy &light &phys\\
$m_\pi$ &315 MeV & 226 MeV & 139  MeV\\
\hline
    \addlinespace[0.4em]
\multicolumn{1}{c}{$m_\pi a_0$~~~}&
~$-0.1673^{+0.0177}_{-0.0191}$~&
~$-0.1001^{+0.0067}_{-0.0064}$~&
~$-0.0436^{+0.0013}_{-0.0012}$~\\
    \addlinespace[0.2em]
\end{tabular}
\end{ruledtabular}
\caption{Scattering length determined from the effective-range expansion.}
\label{tab:a0}
\end{table}
The result at the physical pion mass is depicted together with other lattice QCD based estimations in Fig.~\ref{fig:scattering_lengths}. It shows that the extrapolated value is well in agreement with the value extracted from the Roy equations~\cite{Ananthanarayan:2000ht} as well as overlaps with most recent lattice QCD determinations.

The compilation of phase-shifts for the combined fit to light and heavy energy eigenvalues is depicted in Fig.~\ref{fig:I2phase-shifts}. It also contains the lattice results after extrapolating them to the physical point (right panel). Recall that fitting is performed at the level of energy eigenvalues including full information about cross correlations. Overall we note, that while the LECs of the model have sizable statistical uncertainties, the corresponding error bands on phase-shifts are quite narrow.  Clearly, the prediction of the phase-shifts at the physical point depicted in Fig.~\ref{fig:I2phase-shifts} shows a nice agreement with the available experimental data even far beyond the elastic region. This observation has been also noted in Ref.~\cite{Mai:2018djl} where the two and three pion lattice QCD results of the NPLQCD collaboration~\cite{Beane:2007es, Detmold:2008fn} were discussed.

\section{Conclusion}\label{sec:summary}

We have performed a calculation of the pion-pion elastic scattering in the isopsin $I=2$ channel in two-flavor dynamical lattice QCD. One novel feature compared with other $I=2$ studies is the use of elongated lattices which has proven a cost-effective approach in mapping out the momentum dependence in scattering processes. We considered six ensembles (see Table~\ref{table:gwu_lattice}) with elongation up to a factor of 2 in one of the spatial dimensions, and two pion masses at 315 MeV and 226 MeV. Boosting of the $\pi\pi$ system in the elongated direction is also considered for enhanced energy coverage. In each case, we extract multiple low-lying states using the standard variational method.


The data is analyzed simultaneously at both pion masses making use of the inverse amplitude  method. This unitary model matches the chiral $\pi\pi$ scattering amplitude to next-to-leading order and allows, thus, for chiral extrapolation to the physical point in a wide energy range. In the current application of the method, we allow the pion mass and decay constant to vary and include their full correlations with the energy eigenvalues in the corresponding fit. The scattering length extrapolated to the physical point reads $m_\pi{} a_0=-0.0436(13)$. As an additional check we also perform an effective range expansion, which in combination with the perturbative chiral form of the scattering length gives a consistent value of $m_\pi{} a_0=-0.0455(16)$ at the physical point.

A closer analysis of the $\chi^2$ reveals a slight tension in the model description between light and heavy ensembles if the full energy range is fitted. 
Overall, this suggests that more reliable extraction of low-energy parameters requires results at lower pion mass as well. Another possibility, is to include cross channels, $I=0,1$, in the analysis, which might restrict these constants more strongly. In any case, we note that the phase-shifts extrapolated to the physical point match the experimental phase-shifts. 

Overall our results demonstrate the efficacy of using elongated lattices and IAM analysis as an effective tool in studying hadron-hadron scattering processes from first principles. Work is under way to carry out a global IAM analysis across all three isospin channels of $\pi\pi$ scattering on elongated lattices. Such a study can pave the way for extending to systems that include three pions above threshold.

\begin{acknowledgments}
CC, AA, and FXL are supported in part by the National Science Foundation CAREER grant PHY-1151648 and by U.S. DOE Grant No. DE-FG02-95ER40907. MD and MM  acknowledge  support  by  the  National  Science  Foundation  CAREER grant PHY-1452055. MD is supported in part  by  the U.S.  Department  of  Energy, Office  of  Science,  Office  of  Nuclear  Physics  under  contract  DE-AC05-06OR23177  and  grant  de-sc001658MM. AA gratefully acknowledges the hospitality of the Physics Department at the University of Maryland where part of this work was carried out.
The computations were performed on the GWU Colonial One computer cluster and the GWU IMPACT collaboration clusters. 

\end{acknowledgments}

\bibliography{ALL-REF.bib}

\appendix
\begin{widetext}
\newpage

\section{Extracted energies and correlation matrices}
\label{appendix:fitting}
In this section we summarize results from fitting the correlation function to extract the finite volume energies.  The energies from non-boosted systems were fit with the functional form $A_1 e^{-E_1 t} + A_1' e^{-E_1' t} + B$, while boosted systems were fit with $A_1e^{-E_1t}+A_1'e^{-E_1't}+Be^{-\Delta E t}$ with $\Delta E=\sqrt{m_\pi^2+[2\pi/(\eta L)]^2}-m_\pi$.  The results of all extractions are reported in Table~\ref{table:fit_results}.  The fit window was chosen to minimize the $\chi^2$ per degree of freedom.  The parameter Q is the confidence level of the fit corresponding to the probability that $\chi^2$ is larger then the fit result.  The value of $am_{\pi}$ is extracted from the two point correlation function.  We also include tables of the cross-correlations between the extracted energies for all $\pi\pi$ scattering channels in Table~\ref{tab:cov_isospin2}.

\begin{table}[b]
\renewcommand{\arraystretch}{0.2}
\footnotesize{
\begin{tabular}{@{}*{6}{>{$}c<{$}}*{2}{>{\hspace{1mm}}l!{\hspace{1mm}}}*{1}{>{$0.}c<{$}}@{}} 

\multicolumn{9}{c}{$M_\pi=315\MeV$}\\[0.1cm]
\toprule
\mathbf{P} & \eta & am_{\pi} & n & t_0 & \text{fit window} & aE & $\chi^2$/dof & \multicolumn{1}{c}{Q}\\
\midrule
(0,0,0) &   1.0     &   0.1931(4)   &   1   &   3   &   5-15    &   0.391(1)   &   1.36    &   23    \\
        &           &               &   2   &   3   &   7-17    &   0.669(2)    &   1.24    &   28    \\  
        &   1.25    &   0.1944(3)   &   1   &   3   &   4-13    &   0.393(1)   &   1.23    &   29    \\  
        &           &               &   2   &   3   &   6-14    &   0.577(1)   &   1.10    &   36    \\
        &           &               &   3   &   3   &   5-13    &   0.663(1)    &   0.98    &   42    \\
        &   2.0     &   0.1932(3)   &   1   &   3   &   4-13    &   0.3883(6)   &   0.90    &   48    \\
        &           &               &   2   &   3   &   6-13    &   0.4712(7)   &   0.76    &   52    \\  
        &           &               &   3   &   3   &   2-12    &   0.6509(6)   &   0.97    &   44    \\
        &           &               &   4   &   3   &   5-11    &   0.6579(8)   &   0.76    &   47    \\
        &           &               &   5   &   3   &   5-12    &   0.715(1)   &   0.65    &   58    \\
\midrule
(0,0,1) &   1.0     &   0.1931(4)   &   1   &   3   &   6-13    &   0.526(1)    &   1.02    &   38   \\
        &           &               &   2   &   3   &   6-15    &   0.762(4)    &   1.00    &   42  \\
        &   1.25    &   0.1944(3)   &   1   &   3   &   4-10    &   0.487(1)   &   0.91    &   40  \\
        &           &               &   2   &   3   &   6-14    &   0.729(1)    &   0.61    &   66  \\
        &           &               &   3   &   3   &   4-16    &   0.754(2)     &  1.40    &   19  \\
        &   2.0     &   0.1932(3)   &   1   &   3   &   3-12    &   0.4311(6)   &   1.06    &   38  \\
        &           &               &   2   &   3   &   6-12    &   0.5630(8)   &   0.51    &   60  \\
        &           &               &   3   &   3   &   3-15    &   0.6880(7)   &   1.04    &   40  \\
        &           &               &   4   &   3   &   2-16    &   0.764(1)     &  0.68    &   74  \\
        &           &               &   5   &   3   &   4-10    &   0.782(1)    &  1.01    &   36  \\
\bottomrule
\end{tabular}}
~
\begin{tabular}{@{}*{6}{>{$}c<{$}}*{2}{>{\hspace{1mm}}l!{\hspace{1mm}}}*{1}{>{$0.}c<{$}}@{}}
\multicolumn{9}{c}{$M_\pi=226\MeV$}\\[0.1cm]
\toprule
\mathbf{P} & \eta & am_{\pi} & n & t_0 & \text{fit window} & aE & $\chi/\text{dof}$ & \multicolumn{1}{c}{Q}\\
\midrule
(0,0,0) &   1.0     &   0.1378(6)   &   1   &   3   &   4-12    &   0.280(1)    &   0.84    &   50  \\
        &   1.17    &   0.1374(5)   &   1   &   3   &   5-13   &   0.280(1)    &   0.88    &   47  \\
        &           &               &   2   &   3   &   3-12    &   0.537(1)    &   0.78    &   56  \\  
        &   1.33    &   0.1380(5)   &   1   &   3   &   6-13    &   0.280(1)    &   1.32    &   27  \\
        &           &               &   2   &   3   &   3-12    &   0.490(1)    &   0.76    &   58  \\ 
\midrule
(0,0,1) &   1.0     &   0.1378(6)   &   1   &   3   &   3-12    &   0.446(1)    &   1.10    &   35  \\
        &   1.17    &   0.1374(5)   &   1   &   3   &   3-10    &   0.411(1)    &   1.02    &   38  \\
        &   1.33    &   0.1380(5)   &   1   &   3   &   3-9     &   0.385(1)   &   1.05    &   35  \\
\bottomrule\\[3.35cm]
\end{tabular}

\caption{Energy levels in isospin 2 with fitting details.  $\eta$ is the elongation and $t_0$ is the variational time.}
\label{table:fit_results}
\end{table}

\begin{table*}[tbh!]
\begin{tabular}{  l l | l l  }
&
$
{\rm Cov}[{\mathcal{E}_1}]=
\left(
\begin{array}{cccc}
  8.48   &    7.61   &    7.77   &    5.35   \\
  7.61   &    66.70   &    10.06   &    29.03   \\
  7.77   &    10.06   &    13.26   &    11.71   \\
  5.35   &    29.03   &    11.71   &    122.43   \\
\end{array}
\right)\times10^{-7}$&
&
$
{\rm Cov}[{\mathcal{E}_4}]=
\left(
    \begin{array}{cc}
  15.76   &    7.08   \\
  7.08   &    18.64   \\
    \end{array}
\right)
\times10^{-7}$
\\
&&&\\
&
$
{\rm Cov}[{\mathcal{E}_2}]=
\left(
    \begin{array}{cccccc}
  5.08   &    4.09   &    3.75   &    4.59   &    2.97   &    3.01   \\
  4.09   &    9.55   &    4.55   &    5.07   &    6.43   &    8.41   \\
  3.75   &    4.55   &    12.33   &    4.28   &    9.38   &    5.19   \
\\
  4.59   &    5.07   &    4.28   &    6.79   &    4.04   &    5.91   \\
  2.97   &    6.43   &    9.38   &    4.04   &    14.31   &    10.16  \
 \\
  3.01   &    8.41   &    5.19   &    5.91   &    10.16   &    26.03  \
 \\
\end{array}
\right)\times10^{-7}$
&&
$
{\rm Cov}[{\mathcal{E}_5}]=
\left(
\begin{array}{ccc}
  13.53   &    6.29   &    8.20   \\
  6.29   &    11.70   &    9.13   \\
  8.20   &    9.13   &    11.87   \\
\end{array}
\right)\times10^{-7}$\\
&&&\\
&
$
{\rm Cov}[{\mathcal{E}_3}]=
\left(
    \begin{array}{cccccccccc}
  4.11   &    3.66   &    2.83   &    3.05   &    2.56   &    3.79   \
&    3.39   &    2.50   &    2.92   &    2.84   \\
  3.66   &    5.18   &    3.32   &    3.59   &    3.39   &    3.47   \
&    4.66   &    2.58   &    3.42   &    2.77   \\
  2.83   &    3.32   &    4.10   &    3.57   &    3.45   &    3.12   \
&    4.13   &    3.17   &    5.35   &    4.12   \\
  3.05   &    3.59   &    3.57   &    7.02   &    6.69   &    3.02   \
&    4.65   &    4.09   &    3.75   &    5.97   \\
  2.56   &    3.39   &    3.45   &    6.69   &    11.18   &    2.59   \
&    3.89   &    5.17   &    3.22   &    7.25   \\
  3.79   &    3.47   &    3.12   &    3.02   &    2.59   &    3.84   \
&    3.50   &    2.65   &    3.45   &    3.12   \\
  3.39   &    4.66   &    4.13   &    4.65   &    3.89   &    3.50   \
&    6.11   &    2.95   &    4.82   &    4.11   \\
  2.50   &    2.58   &    3.17   &    4.09   &    5.17   &    2.65   \
&    2.95   &    5.03   &    2.91   &    5.08   \\
  2.92   &    3.42   &    5.35   &    3.75   &    3.22   &    3.45   \
&    4.82   &    2.91   &    10.26   &    5.68   \\
  2.84   &    2.77   &    4.12   &    5.97   &    7.25   &    3.12   \
&    4.11   &    5.08   &    5.68   &    17.59   \\
    \end{array}
\right)
\times10^{-7}$
&&
$
{\rm Cov}[{\mathcal{E}_6}]=
\left(
\begin{array}{ccc}
  13.28   &    4.73   &    7.67   \\
  4.73   &    10.59   &    6.93   \\
  7.67   &    6.93   &    11.71   \\
\end{array}
\right)
\times10^{-7}$\\
\end{tabular}
\caption{Covariance matrices for each ensemble. The ordering of the levels is consistent with the order in Table~\ref{table:fit_results}.
}
\label{tab:cov_isospin2}
\end{table*}

\end{widetext}
\end{document}